# A Proposed Sub-optimal Power Allocation using Simulated Annealing in Cognitive Radio Networks

## Abd Elhamed M. Dawoud[1], Mona Shokair[2], Mohamed Elkordy[3] and Said ElHalafawy[4]

*[1,2,3,4] Department of Electrical and Electronics Engineering, Faculty of Electronic Engineering,
Menofia University, Menouf, Egypt*

**Abstract:** Due to the rapid demand for wireless services and the increase in the wireless device count, there is a lack of available spectrum bands which constrain the further development of wireless communication .Therefore, Cognitive Radio (CR) has been adopted as a promising solution because of its ability to exploit the inefficiently used spectrum of licensed bands. Orthogonal Frequency Division Multiplexing (OFDM) become the enabling technique for CR due to its flexibility of allocating the available spectrum in dynamic environment. In this paper, a proposed distributed resource allocation framework based on Simulated Annealing (SA) algorithm for downlink OFDM-Based Cognitive Radio Network (CRN) will be applied. This algorithm gives less computational complexity for maximizing the total SUs transmission capacity. Moreover, the interference introduced from other Secondary Users (SUs) will be considered. For the sake of comparison, Lagrange dual method will be used. Simulation results showed that the proposed algorithm gives a better transmission capacity compared with Lagrange dual method. The parameters which are considered for comparison are maximum transmitted power, number of Primary Users (PUs) and number of SUs.

**Keywords:** Cognitive radio, OFDM, Power allocation, Simulated Annealing (SA), evolutionary Algorithms, distributed algorithm.

## 1. INTRODUCTION

Along with the rapid growing of wireless communication development, a more services are required and with a more wireless devices are used due to wireless communication technology, the licensed spectrum bands are becoming increasingly congested. According to FCC [1], the licensed bands are inefficiently allocated all the time. CRN has been recognized as a promising network for efficient spectrum utilization. It has the ability to sense the unused bands and dynamically allocate them to unlicensed users. In this network, SU has the ability to exploit Primary User (PU) band as long as the interference introduced to the PU does not exceed the accepted level [2].

Actually, OFDM has been adopted as a transmission technique for CR system as OFDM has the ability of flexible spectrum allocation and underlying sensing due to FFT module in its receiver. Due to OFDM orthogonal property, there is no introduced interference to the adjacent channel. However, CR is a heterogeneous communication system in which there may be some energy leakages of OFDM signal to adjacent channel which yields mutual interference between PU and SU. The amount of interference between PU and SU's sub-carriers depends on the power allocated in that sub-bands and the spectral distance between them [3-4]. The problem of power allocation for OFDM-based CRN has been studied in literature. In [5], a survey on different resource allocation scheme for OFDM multiuser Downlink system was introduced. Optimal and sub-optimal power techniques had been investigated in [6]. In which the authors had been developed an optimal power allocation algorithm that maximizes the downlink transmission capacity of the CR user while the interference introduced to the PU remains within a certain level. They had also proposed two suboptimal power allocation algorithms with less complexity but can achieve performance close to the optimal one. Ref [7], defined an optimal and two sub-optimal power allocation techniques to reduce the complexity of the optimal one. The authors claimed that the proposed algorithms are better than the classical water-filling algorithm and uniform power loading Algorithm. Two sub-optimal algorithms has considered in Ref [8], which are fair in the sense that they try to allocate bits to users who have not received their fair share of service as much as possible. In fact, many researchers have focused on using evolutionary algorithms such as Genetic algorithms (e.g. [9], [10]) and Particle Swarm Optimization for optimization in wireless communication. Memetic Algorithm (MA), a modified Genetic algorithm, was proposed as a fast and efficient technique for dynamically allocating subcarriers, transmit powers and bits to secondary users in [11]. Particle swarm optimization technique for adaptive power allocation technique for multi-user downlink OFDM CR had been investigated in [12].

In this paper, the problem of power allocation for OFDM-based CRNs will be investigated, and a proposal for computationally efficient sub-optimal power loading algorithm that can maximize SUs transmission capacity while satisfying sub-channel, total transmit power constraint and the interference threshold introduced to PUs. The interference from PUs and other SUs were taken into account. The proposed Algorithm is based on Simulated Annealing (SA) optimization technique. This scheme can be used in a centralized or distributed manner. Comparison between our proposed algorithm and other algorithms will be made. Moreover, the effect of some parameters such as transmitted power, number of PUs and SUs will be made.

The rest of this paper is organized as follows. In Section II, system model will be described. Problem formulation and proposed scheme will be implemented in Section III, while simulation results will be introduced in Section IV, followed by conclusions in Section V.

## 2. SYSTEM MODEL

In this paper, an ad-hoc downlink (CRN) is assumed as shown in Fig.1. Where PUs are served by the Primary User Base Station (PU-BS) and SUs can either occupy PU available spectrum bands or the adjacent spectrum of PU. OFDM technique involves SUs communication where OFDM divides the available bands into sub-bands with a flat fading. In Fig. 2, the spectrum domain model of the system model is illustrated where $B_p$ represents the bandwidth of PU which may be not fully utilized by PU. Therefore, SU can exploit that spectrum. $B_s$ represents the bandwidth that is divided by OFDM for SUs communication. SUs have full knowledge of the aggregated interference from the PU. Actually, there are three channel gains are known in our model. These are $h_k^{sp}$, $h_k^{ps}$ and $h_k^{ss}$ terms which represent the channel gain from the SU transmitters to PU receiver, channel gain from PU transmitters to SU receiver and channel gain between SU transmitters and SU receiver pairs on subcarrier k, respectively. All of these channel gains are assumed to be known to each one of SUs. As CRN is a heterogeneous communication network there would be a mutual interference between PUs and SUs pairs and between SU pairs. In this paper, two kinds of interference will be taken into account, the interference introduced between SUs pairs and the interference from PUs to SUs.

### A. Interference From SU to PU

Assuming that the transmitted signal on the subcarrier k is an ideal Nyquist pulse, $T_s$ is OFDM symbol duration and $p_k$ is the total transmitted power on subcarrier k. Therefore, the power spectral density of OFDM signal of $k^{th}$ subcarrier according to [3] is given by:

$$\emptyset_k(f) = p_k T_s \left( \frac{\sin \pi f T_s}{\pi f T_s} \right)^2 \qquad (1)$$

Where the interference introduced to PU by SU can be expressed as:

$$I_k(p_k) = p_k |h^{sp}|^2 T_s \int_{d_k - B_p/2}^{d_k + B_p/2} \left( \frac{\sin \pi f T_s}{\pi f T_s} \right)^2 df \qquad (2)$$

Where $d_k$ is the spectral distance between the subcarrier $k$ and the central Frequency of $B_p$ .

### B. Interference from PU to SU.

Suppose that $\varphi_{pu}(e^{j\omega})$ is the PSD of PU signal with an amplitude of $p_{pu}$ . According to [6], the interference introduced by PU to the $k^{th}$ subcarrier can be written as:

$$J_k(p_{pu}) = \int_{d_k - B_s/2}^{d_k + B_s/2} |h_k^{ps}|^2 \epsilon\{I_k(\omega)\} d\omega \qquad (3)$$

Where $\epsilon\{I_K(\omega)\}$ is the PSD of PU signal after N-point Fast Fourier Transform (FFT), It can be expressed as [3],

$$\epsilon\{I_K(\omega)\} = \frac{1}{2\pi N} \int_{-\pi}^{\pi} \varphi_{pu}(e^{j\omega}) \left( \frac{\sin(\omega - \varphi)N/2}{\sin(\omega - \varphi)/2} \right)^2 d\varphi \qquad (4)$$

## 3. PROBLEM FORMULATION AND PROPOSED SCHEME

### A. Problem formulation

The problem of resource allocation, is to intelligently allocate limited power and bandwidth resources among users according to each user requirements. It had been discussed heavily from the point of view of power and bit allocation [8]. Through formulating it as an objective problem and aiming to maximize the SU's transmission capacity through optimization methods. Taking into account the constraints of total SUs transmitted power, the interference introduced to PU not to exceed the allowed level and the interference generated from other SUs.

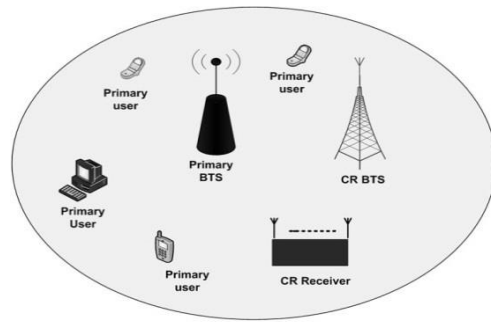

Figure 1. System model of an ad-hoc cognitive radio.

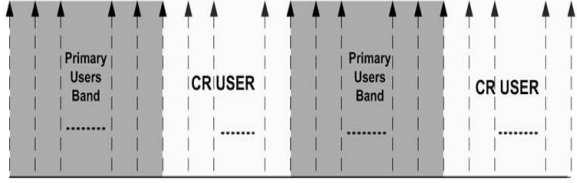

Figure 2. The spectral domain model for cognitive radio.

Considering CRN capacity as the objective function and from Shannon theorem the bit rate of the $m^{th}$ SU on the $k^{th}$ subcarrier, assuming that every SU can occupy only one sub-channel at one time, can be written as:

$$R_{k,m}(p_k) = B_s \, log_2\left(1 + \left(\frac{|h_k^{ss}|^2 p_{k,m}}{\sigma^2 + \sum_{l=1}^{L} J_k^{(l)} + \sum_{i \neq k} IN_k}\right)\right)$$
$$1 \leq k \leq K \qquad (5)$$

Where $\sigma^2$ is Noise variance, the second part in the denominator represents the interference from the PUs and the third part expresses the interference of other SUs that can be denoted by

$$IN_k(f) = p_k |h_k^{ss}|^2 T_s \int_{d_k - B_s/2}^{d_k + B_s/2} \left(\frac{\sin \pi f T_s}{\pi f T_s}\right)^2 df \qquad (6)$$

The goal of this paper is to maximize the transmission rate of SUs that can be defined by,

$$C = max \quad _{P_k} \, log_2\left(1 + \left(\frac{|h_k^{ss}|^2 p_k}{\sigma^2 + \sum_{l=1}^{L} J_k^{(l)} + \sum_{i \neq k} IN_k}\right)\right)$$
$$(7)$$

Subject to:

$$\sum_{k=1}^{K} p_k \leq p_{max} \qquad (8)$$

$$\sum_{k=1}^{K} I_k^l(p_k) \leq I_{th} \qquad (9)$$

$$p_k \geq 0 \qquad k = 1,2,\dots,K \qquad (10)$$

$C$ represents system's transmission capacity, $B_s$ is bandwidth of each SU Subcarrier, $p_{max}$ is the maximum transmitted power allowed for all of SUs and $I_{th}$ threshold interference prescribed by PU. $K$ is the number of subcarriers and $p_{k,m}$ represent the allocated power for $m^{th}$ SU at the $k^{th}$ subcarriers.

According to [13], the optimization problem is convex. The optimal solution can be found using the dual Lagrange method as follow. By defining $\mu$ and $\lambda$ as Lagrange dual multipliers, the Lagrange equation can be expressed as:

$$\mathcal{L}(p_k, \mu, \lambda) = log_2\left(1 + \left(\frac{|h_k^{ss}|^2 p_k}{\sigma^2 + \sum_{l=1}^{L} J_k^{(l)} + \sum_{i \neq k} IN_k}\right)\right) -$$
$$\mu\left(\sum_{k=1}^{K} p_k - p_{max}\right) - \lambda\left(\sum_{k=1}^{K} I_k^l(p_k) \, I_{th}\right) \qquad (11)$$

By applying the partial derivatives and equalling to zero, the result can be defined as,

$$\frac{\partial \mathcal{L}(p_k, \mu, \lambda)}{\partial p_k} = \frac{|h_k^{ss}|^2}{(\sigma^2 + \sum_{l=1}^{L} J_k^{(l)} + \sum_{i \neq k} IN_k + |h_k^{ss}|^2 p_k) \, ln \, 2} - \mu - \lambda = 0, \qquad (12)$$

$$\frac{\partial \mathcal{L}(p_k, \mu, \lambda)}{\partial \mu} = \sum_{k=1}^{K} p_k - p_{max} = 0, \qquad (13)$$

$$\frac{\partial \mathcal{L}(p_k, \mu, \lambda)}{\partial \lambda} = \sum_{k=1}^{K} I_k^l(p_k) - I_{th} = 0. \qquad (14)$$

To find the optimal solution, these equations should be solved firstly by obtaining the values of $\mu$ and $\lambda$. The Lagrange multipliers values can be found by using an exhaustive search algorithm and the optimal power $p^*$, which gives the best value of the transmission capacity that can be written as:

$$p^* = max\left\{0, \frac{1}{(\lambda + \mu) \, ln \, 2} - \frac{\sigma^2 + \sum_{l=1}^{L} J_k^{(l)} + \sum_{i \neq k} IN_k}{|h_k^{ss}|^2}\right\} \qquad (15)$$

The dual Lagrange method has a high computational complexity of $O(K^2 \log_2 K)$. The objective of this paper is to introduce an efficient sub-optimal power allocation technique based on Simulated Annealing (SA). The proposed scheme reduces the complexity compared to the dual Lagrange algorithm and also increases system performance. It is also can be implemented in a distributed manner by SUs.

### B. Proposed scheme

In fact, SA is a heuristic evolutionary probabilistic algorithm that was proposed by Kirkpatrick, S., Gelatt, C.D., and Vecchi, M.P in [14], to find the global minimum of the optimization problem without being easily fooled with local minima. It is an efficient and a powerful optimization technique that had been used to solve optimization problem in many fields such as real life problems. SA imitates the annealing process used in Metallurgy and draws an analogy between the cooling of a material (search for minimum energy state which corresponds the minimum value of the objective function), then solving of an optimization problem will be fulfill. Annealing, which is known as a thermal process for obtaining low energy states of substance in heat bath, consist of two steps. Firstly, heating up the solid until it melts. Secondly, slowly cooled down in a control manner until it solids back. The final properties of this substance depend strongly on the cooling schedule applied; if it cools down quickly the resulting substance will be easily broken due to an imperfect crystal structure, if it cools down slowly the resulting structure will be well organized and strong. When using metropolis algorithm for annealing simulation, the control parameter is the temperature. The temperature is used to define how and when the new solution is accepted or rejected.

Supposing that $E_i$ is the energy of state i, $E_j$ is the energy of state j and $E_i - E_j$ is the energy difference between state i and state j. If the energy difference is less than or equal to zero, $E_j$ will be accepted as the best state solution. Otherwise, it still has an opportunity to be the best state solution with a probability equal to [14],

$$\rho = exp\left(\frac{E_i - E_j}{K_B T}\right) \qquad (16)$$

Where $K_B$ the Boltzmann constant and T is the temperature in the heat path, which represent power in this case. Although, the new energy state does not give a better value than the old one, the algorithm accepts it with a certain probability and this is the key behind the ability of SA to not be trapped in the local minima. As the algorithm proceeds and T decreases over iterations, the probability of being at high energy state will be less and it is difficult to accept that point.

The proposed algorithm can be summarized in the following steps.

- Choose a random power vector $p_i$, select a high initial system power P, a stopping criteria ε, specify the cooling (i.e. annealing) schedule and set t = 0.
- Evaluate $E(P_i)$ using a simulation model where E represents objective function value.
- Perturb $P_i$ to obtain a neighbouring power vector $(P_{i+1})$.
- Evaluate $E(P_{i+1})$ using a simulation model.
- If $E(P_{i+1}) < E(P_i)$, $P_{i+1}$ is the new current solution and set t = t+1.
- Else if $E(P_{i+1}) > E(P_i)$, create a random number r in the range (0, 1), then accept $P_{i+1}$ as the new current solution with a Probability $e^{\left(\frac{-\Delta}{P}\right)} \geqslant r$ Where $\Delta = E(P_{i+1}) - E(P_i)$ and set t = t+1. Else, go to step (3).
- If $|P_{i+1} - P_i| < \varepsilon$ , P is small, and satisfying the constraints STOP. Else, go to step (8).
- Reduce the system power according to the cooling schedule.
- Terminate the algorithm if the iteration time reaches the maximum value. Display the optimal power vector and the corresponding best objective function value that represents the system's transmission capacity.

As mentioned before, the complexity of the optimal Dual Lagrange algorithm is $O(K^2 \log_2 K)$ which is very high and getting the optimal value P* depends on the values of $\mu$ and $\lambda$ . On contrary, SA has only a complexity of $O(\log_2 K)$. It is a simple algorithm, derivative free and easy to be realized.

## 4. SIMULATION RESULTS

In this Section, the performance evaluations of SA as the proposed technique will be discussed based on numerical simulation results. For an ad hoc CRN with 2 PUs, 8 SUs and AWGN of zero mean and noise variance $\sigma^2$ equals to $10^{-6}$. The channel gains are assumed to be randomly generated by Rayleigh distribution and total system BW is 12.8 MHZ which is divided into 32 subcarriers of 0.4 MHZ for each one. The powers of PUs are assumed to be the same and equal to 0.01W and the threshold interference of the two PU is 1mW. Setting initial power equal to 100W, stopping criteria ε =$10^{-6}$ and exponential annealing scheduling equals to P *$0.95^t$. In Fig. 3, the system's capacity evaluation of the Simulated Annealing as the proposed scheme compared to the dual Lagrange method had been studied. Which show that the proposed scheme gives a better performance of about 5% increasing in the system's transmission capacity at the saturation and about 35% at the low power.

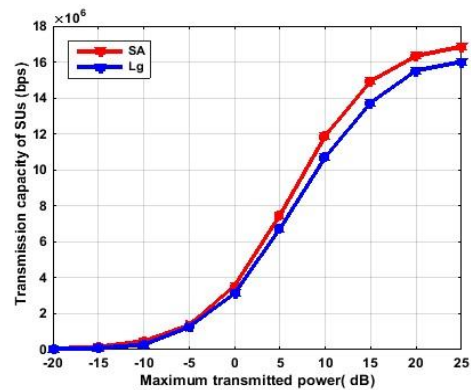

Figure 3. The system performance of Dual Lagrange method and Simulated Annealing.

Total transmission capacity of SUs versus iteration index of SA at $p_{max}$ = 5dBw is depicted in Fig.4. The power allocated to each subcarrier at maximum transmitted power of 5dBw is illustrated in Fig.5

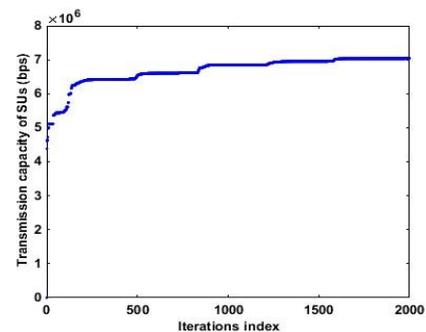

Figure 4. Total throughput of SUs versus iteration index at $p\_max$ = 5dBw.

Considering that sub-carrier 1 and 2 are allocated to PU. Therefore, the power allocated to each one is so small compared to other sub-carriers. In Fig. 6, the effect of increasing the number of PU on the system performance had been figured. As shown from this figure, at power values higher than -5dBW, if the number of PU increases, the transmission capacity of SUs goes down and this is due to the fact that the increase in the number of PU raises the amount of interference aggregated to the SU and also a less subcarriers will be available for SUs transmission. But at lower power values, the effect of changing the number of PUs is so small. The effect of the number of SUs on the system's transmission capacity is highlighted in Fig. 7. As the number of the SU increases, the system's transmission capacity improved until a certain power level where the number of SUs does not affect system performance, after that point as the number increases the capacity degrades due to the increased aggregated interference from other SUs.

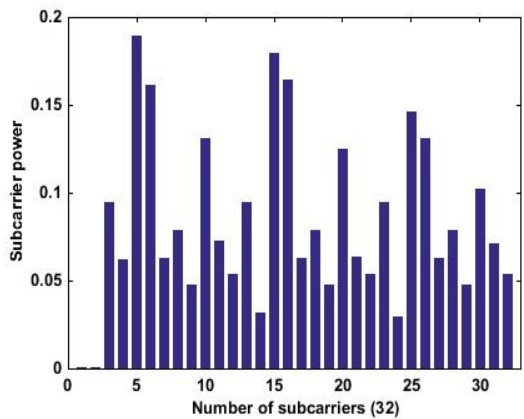

Figure 5. The power allocated to each subcarrier at p_max =5dBw.

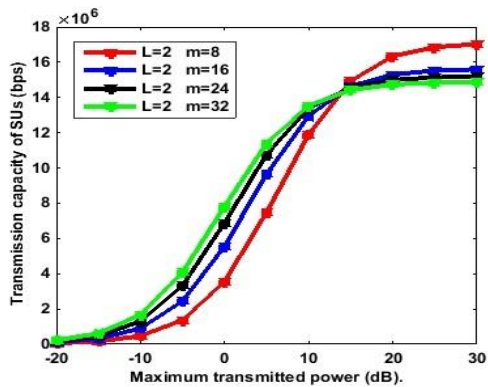

Figure 6. The effect of the number of PUs on the system performance.

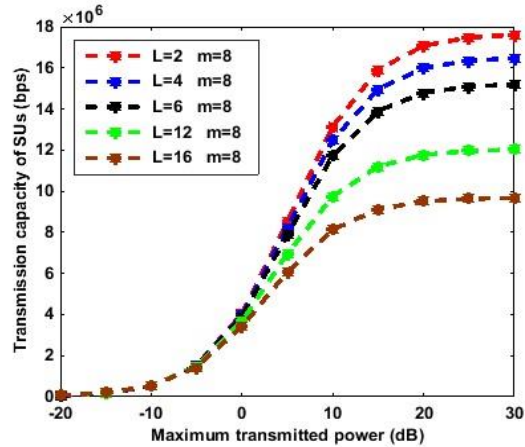

Figure 7. The effect of the number of SUs on the system performance.

## 5. CONCLUSION

Cognitive Radio has been adopted as the promising technique that has the ability to solve the problem of resource allocation in an intelligent way. One of the major challenges to enable CR is to find an optimal power allocation algorithm that can efficiently allocate the power under some constraints, a required quality of service and the method of implementation centralized or distributed. As shown in this paper, a proposed a sub-optimal power allocation algorithm based on Simulated Annealing gives a system performance near the optimal one but with less complexity. The performance of the proposed algorithm had been evaluated under various parameters such as number of PUs and number of SU and compared to the optimal dual Lagrange algorithm. The simulation results showed that the proposed scheme gives a system transmission capacity near to the optimal solution but, with less complexity.

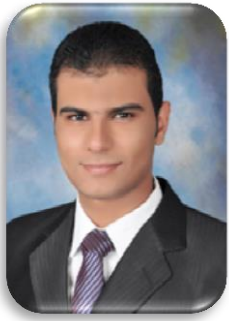

**Abd Elhamed M. Dawoud** received his B.Sc. degree in electronics and electrical communications engineering from the Faculty of Electronic Engineering Menofia University, Egypt, in 2010. Currently, he is demonstrator at the Department of electronics and electrical communications engineering, working towards his M.Sc. degree at the Faculty of Electronic Engineering, Menofia University, Egypt. His current research areas of interest include cognitive radio networks, power control in cognitive radio networks and 5G communication. His research interests include Mobile Communication, Cognitive Radio Network and optimization algorithms.

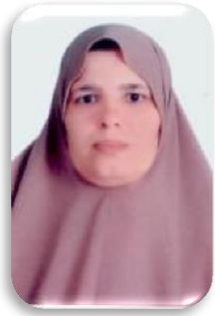

**Mona Shokair** received the B.E. and M.E. degrees in electronics engineering from El Menofia University, El-Menofia, Egypt, in 1993 and 1997, respectively. She received Ph.D. from Kyushu University, Japan, 2005. She was lecturer in El-Menofia University from 2006 to 2011. Since 2012, she is Associated Professor in El-Menofia University. She received VTS chapter IEEE award from Japan in 2003. Now her current research is in OFDM system, WIMAX system and cognitive radios.

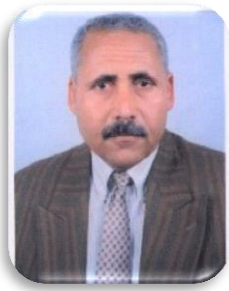

**Mohamed Elkordy** He Received the B.Sc. Excellent Degree with Honors in Communications Engineering from El Menofia University, El-Menofia, Egypt, in 1979, the M.Sc. Degree in Electrical Communications from Menofia University, Egypt, 1985, and the Ph.D. Degree in Electrical Communications from Menofia University, EGYPT, 1991. Now he is interested in Digital signal processing, Ultra Wideband Radars, and Cognitive Radio Networks.

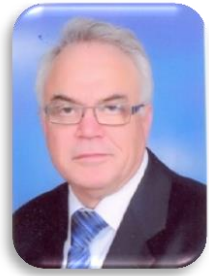

**Said El Halafawy** received his BSc and MSc degrees in Electronic Engineering from the Menofia University, Menouf, Egypt, in 1978 and 1984 respectively. and the PhD from Dept. of Electrical Engineering, Faculty of Electrical and Mechanical Engineering, Plzen, Czech Republic in 1990. He joined the teaching staff of the Department of Electronics and Electrical Communications, Faculty of Electronic Engineering, Menofia University, Menouf, Egypt, in 1991. His current research areas of interest include wireless communication, antennas, microwaves, numerical techniques, image processing.